\documentclass[pageno]{jpaper}

\usepackage[normalem]{ulem}

\usepackage{mathptmx} 
\usepackage[ruled,vlined]{algorithm2e}

\usepackage{amsmath}
\usepackage{amssymb}
\usepackage[]{algorithm2e}
\usepackage{algorithmicx}
\usepackage{graphicx}
\usepackage{caption}
\usepackage{breqn}
\usepackage{authblk}

\usepackage{soul}
\usepackage{letltxmacro}
\usepackage{hyperref}
\LetLtxMacro{\oldhl}{\hl}

\newcommand{\ignore}[1]{}
\newcommand{\DESIGN}{E-ORAM\xspace}

\title{\textbf{Egalitarian ORAM: Wear-Leveling for ORAM}}

\ignore{\author{Yi Zheng$^1$ \and Aasheesh Kolli$^{1,2}$ \and Shaizeen Aga$^3$
\vspace{-1em}
}
\date{%
    $^1$Pennsylvania State University\\%
    \{ yfz5026, akolli \} @psu.edu\\
    $^2$Google, Inc.\\%
    $^3$Advanced Micro Devices, Inc.\\
    shaizeen.aga@amd.com
}
}

\author{{Yi Zheng\textsuperscript{1}, Aasheesh Kolli \textsuperscript{1,2}, and Shaizeen Aga \textsuperscript{3}\endgraf 
\itshape 
\textsuperscript{1}Pennsylvania State University \quad 
\textsuperscript{2}Google, Inc. \quad
\textsuperscript{3}Advanced Micro Devices, Inc. \endgraf 
\textsuperscript{1}  \{ yfz5026, akolli \} @psu.edu \quad 
\textsuperscript{3}shaizeen.aga@amd.com
}
}
\date{}

\begin{document}

\maketitle




\vspace{-1em}
\begin{abstract}

While   {non-volatile memories} ({NVMs}) provide several desirable characteristics like better density and   {comparable} energy efficiency than DRAM, DRAM-like performance, and disk-like durability, the limited endurance   {NVMs} manifest remains a challenge with these memories. Indeed, the endurance constraints of   {NVMs} can prevent solutions that are commonly employed for other mainstream memories like DRAM from being carried over as-is to   {NVMs}. Specifically, in this work we observe that, Oblivious RAM (ORAM) primitive, the state-of-art solution to tackle memory bus side channel vulnerability, while widely studied for DRAMs, is particularly challenging to implement as-is for {NVMs} as it severely affects endurance of   {NVMs}. This is so, as the inherent nature of ORAM primitive causes an order of magnitude increase in write traffic and furthermore, causes some regions of memory to be written far more often than others. This non-uniform write traffic as manifested by ORAM primitive stands to severely affect the lifetime of   {non-volatile memories} (1\% of baseline without ORAM) to even make it impractical to address this security vulnerability. 

To address this challenge, in this work, we propose \textbf{Egalitarian ORAM (\DESIGN)}, which discusses an ORAM design for {NVMs} while achieving close to ideal lifetime of   {non-volatile memory}. To do so, we observe that the inherent nature of ORAM primitive can be exploited to design a wear-levelling algorithm which spreads the additional write traffic generated by ORAM more uniformly across the entire   {non-volatile memory} space. We demonstrate that in comparison to existing state-of-art wear-levelling algorithm which in presence of ORAM can only attain 2.8\% lifetime, proposed \DESIGN design attains approximately 91\% lifetime allowing it to perform orders of magnitude more reads/writes before system failure while introducing less than 0.02\% performance overhead.

\end{abstract}


\section{Introduction}
\label{sec:intro}

Memory system innovations continue to be fueled by the ever-present and ever-growing demands for increased memory capacity and bandwidth. One such innovation is Non-volatile memories (NVMs) (including PCM~\cite{PCM}, STTRAM~\cite{STTRAM}, and ReRAM~\cite{ReRAM}). These memories offer several desirable characteristics like better density and  comparable energy efficiency than DRAM, DRAM-like performance, and disk-like durability. Consequently,  NVMs potentially stand to play an important role in future memory systems. 

However, deploying  NVMs widely is not devoid of challenges. One of the challenges associated with  NVMs is the limited endurance  NVMs manifest, that is, the physical properties of memory cells in  NVMs dictate a limit on number of writes to the memory cell (typically between $10^7 $ to $10^8$ \cite{pcm_lifetime}). Beyond this limit, the memory cell may lose the ability to change state causing data errors. This in turn can lead to system failure when enough memory lines reach endurance limit. This happens when the number of spare memory lines is lower than memory lines that have incurred more writes than dictated by the endurance limit. 

We observe in this work that the endurance constraints of  NVMs make it impractical to carry-over solutions which are widely studied for mainstream memories like DRAM to  NVMs. Specifically, we observe this to be true in the context of ORAM primitive, which is typically studied to address memory bus side channel vulnerability for DRAM systems. To exploit this vulnerability, an attacker taps the memory bus to learn the memory address trace of the program. Prior work~\cite{msr_access_trace_info_leak} has shown that even in presence of data encryption, an attacker can learn sensitive information about the program simply by observing the memory address trace of the program. To tackle this vulnerability, the ORAM primitive accesses an order of magnitude more memory locations on each memory access and further shuffles memory on each access. As such, under ORAM, any memory address trace is computationally indistinguishable from any other address trace of the same length. 

\begin{figure}
         \centering
         \includegraphics[width=0.8\linewidth]{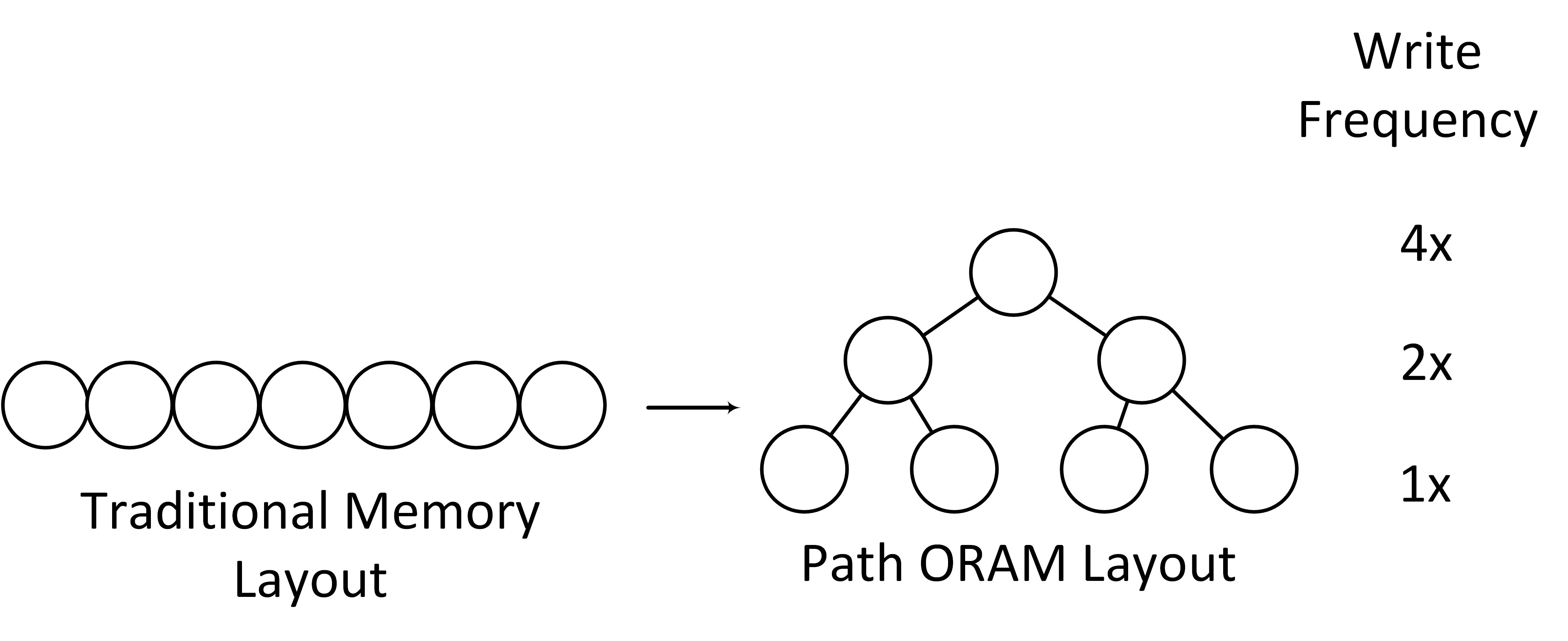}
         \caption{Traditional memory layout (left) vs Path ORAM layout (right). Path ORAM leads to an exponentially distributed memory access pattern. The numbers on the right shows the expected number of writes per node at each level. 
         }
         \label{fig:path_oram_draw_back}
     \end{figure}

However, this very nature of ORAM primitive while effective in addressing memory bus side channel severely affects endurance of  NVMs. This is so, as memory lines are less likely to reach endurance limit in presence of uniform memory write traffic (write traffic uniformly distributed across the entire memory space) as compared to non-uniform memory traffic where few memory lines are written more often than others. The most state-of-art implementation of ORAM, Path-ORAM~\cite{path_oram}, however, manifests exponential write distribution severely affecting  NVM endurance. This happens as Path-ORAM organizes memory as a binary tree and on each memory access reads and writes a entire path in this tree. Fig. \ref{fig:path_oram_draw_back} shows the memory layout of Path ORAM. As shown in Fig. \ref{fig:path_oram_draw_back}, Path ORAM arranges memory as a balanced binary tree structure, where each node is written equally at the same level. From the root to leaf, the expected number of writes to each node decreases exponentially.  This leads to an exponential write distribution, where nodes closer to the root node in the memory tree are written far often than leaf nodes in the tree. This non-uniform write traffic as manifested by ORAM primitive stands to severely affect 
the lifetime of  non-volatile memories (1\% of baseline without ORAM) so as to even make it impractical to address memory bus side channel for  NVMs. 

Note that, prior works~\cite{Start-Gap} have observed that even typical application behavior can exacerbate the endurance constraints of  NVMs as they manifest non-uniformity in write traffic. As such, to address this, several \textit{wear-levelling} algorithms have been proposed by prior works~\cite{age_based_wl, random_swap_wl, Start-Gap, one_percent_lifetime, march_based_wl}. The tenet of these algorithms is to spread application write traffic as uniformly as possible across the  non-volatile memory space by employing memory line remapping as necessary. However, as we show in this work, the ORAM primitive severely stresses these prior solutions and as such, existing state-of-art wear-levelling algorithm~\cite{Start-Gap} in presence of ORAM can only attain 2.8\% lifetime. 

To tackle  NVM endurance constraints while addressing memory bus side channel vulnerability with ORAM primitive, in this work we make the observation that the inherent nature of ORAM primitive can be exploited to design a wear-levelling algorithm which directly tackles the endurance pressure of ORAM primitive. Specifically, while the Path-ORAM~\cite{path_oram} implementation of ORAM primitive by its very nature manifests an exponential write distribution, it does so in a \textit{deterministic} manner. That is, the expected number of writes to a given memory line is statically known depending on where in the Path-ORAM tree the memory line belongs (e.g., root note is written twice as much as its children). We use this information provided by the inherent nature of the ORAM primitive to augment a state-of-art wear-levelling algorithm, \textit{Start-Gap}~\cite{Start-Gap} and make it ORAM primitive aware. We term this ORAM design for  NVMs which directly tackles  NVM endurance constraints, \textbf{\DESIGN}. We show how our proposed design can attain 91\% lifetime
and perform orders of magnitude more reads/writes before system failure while introducing less than 0.02\% performance overhead. 

The contributions of this work are:

\begin{itemize}
    \item We observe in this work that the endurance constraints of  non-volatile memories (NVMs) make it impractical to implement Oblivious-RAM (ORAM) primitive for  NVMs, a state-of-art defense against memory-bus side channel vulnerability, as it leads to a system lifetime of 1\% of baseline system without ORAM. 
    \item To address this challenge, we make the observation that the inherent nature of ORAM primitive can be exploited to design a wear-levelling algorithm which directly tackles the endurance pressure of ORAM primitive.  
    \item We harness the above observation in our proposed ORAM design for  NVMs, termed, \textbf{\DESIGN}, in which we augment a state-of-art wear-levelling algorithm, \textit{Start-Gap}~\cite{Start-Gap} with ORAM primitive awareness. 
    \item We demonstrate that while state-of-art wear-levelling algorithm~\cite{Start-Gap} in presence of ORAM can only attain 2.8\% lifetime, proposed \DESIGN design attains 91\% lifetime allowing it to perform orders of magnitude more reads/writes before system failure while introducing less than 0.02\% performance overhead.
\end{itemize}

\section{Background and Motivation}
\label{sec:background_motivation}

\begin{figure*} 
     \subfloat[Initial Status of the ORAM tree, stash, and PosMap. A read request to block C is issued by the application.]{\includegraphics[width=0.65\columnwidth]{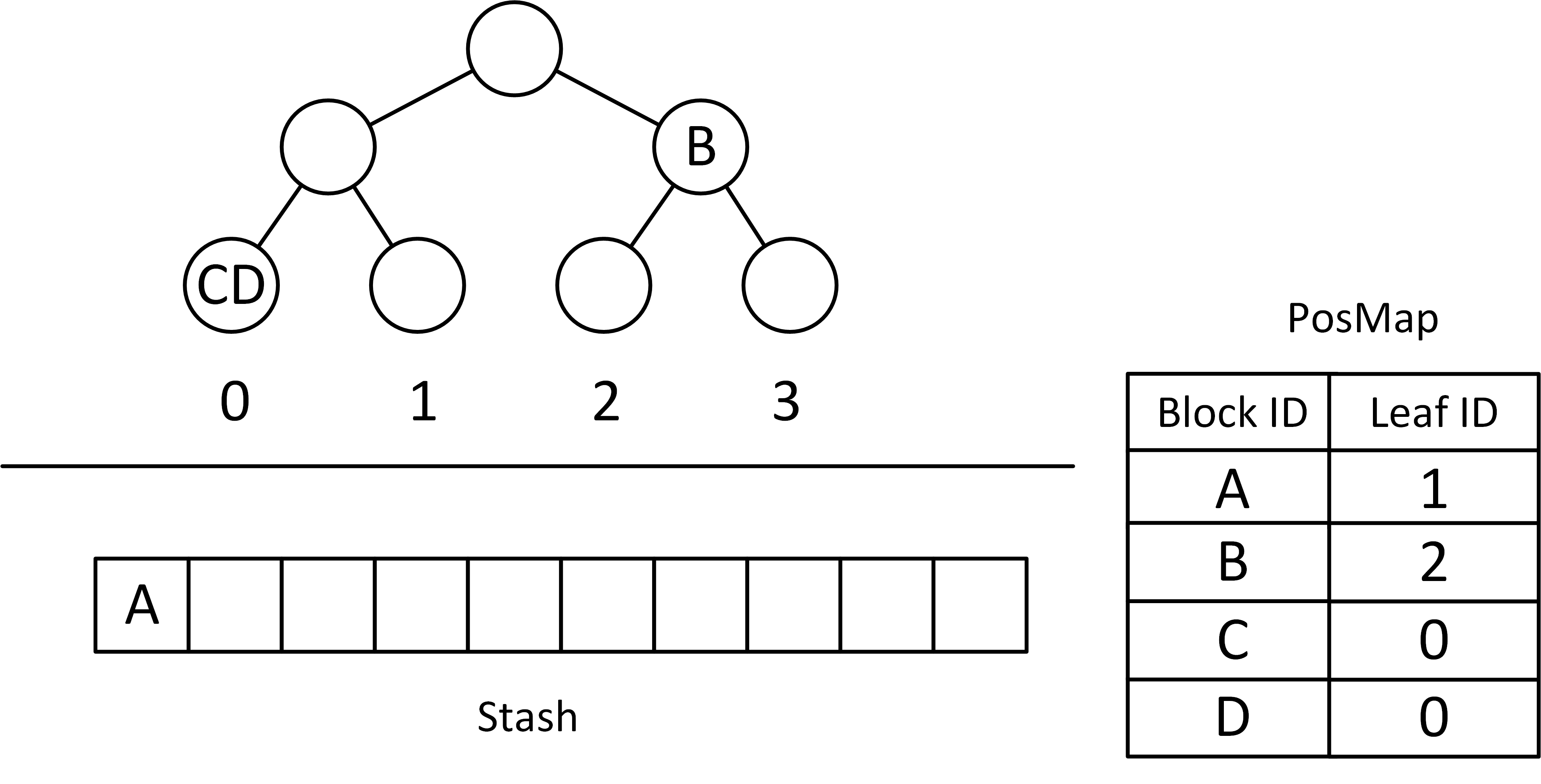}\label{fig:path_oram_example_init}}%
    \hspace{0.5cm}
     \subfloat[Status of the ORAM tree, stash, and PosMap after read accesses to all nodes on the path from root to leaf 0. Block C and D are loaded into the stash. Then, ORAM remaps block C to leaf 1.]{\includegraphics[width=0.65\columnwidth]{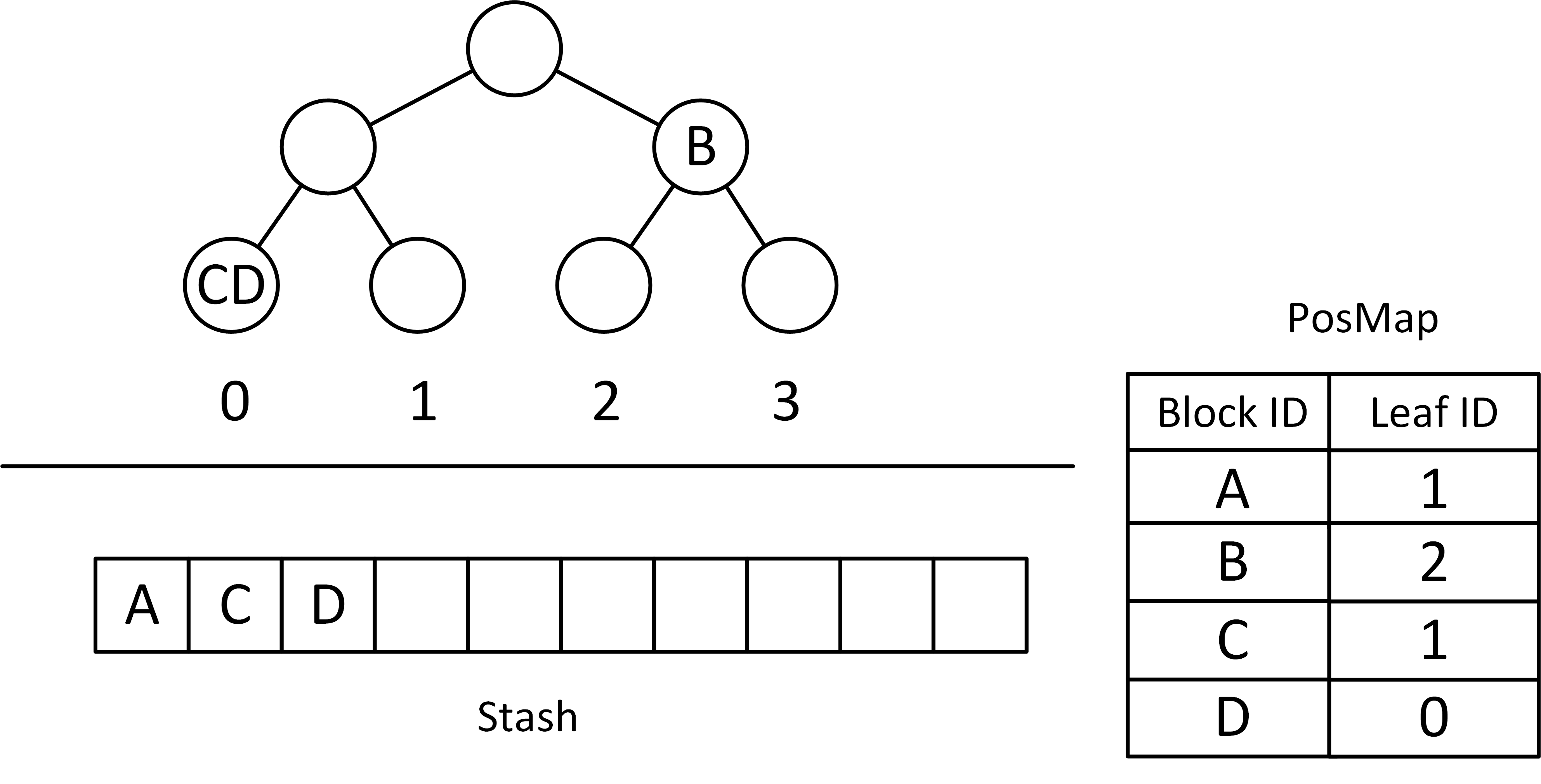}\label{fig:path_oram_example_after_read}}%
    \hspace{0.5cm}
     \subfloat[Status of the ORAM tree, stash, and PosMap after write back operation. The write back path is the same as the read path. Block D is mapped to leaf 0, so it is pushed to leaf 0. Block C is mapped to leaf 1, as such, unlike before it cannot be placed in leaf block. Block A can be drained from stash to the tree.]{\includegraphics[width=0.65\columnwidth]{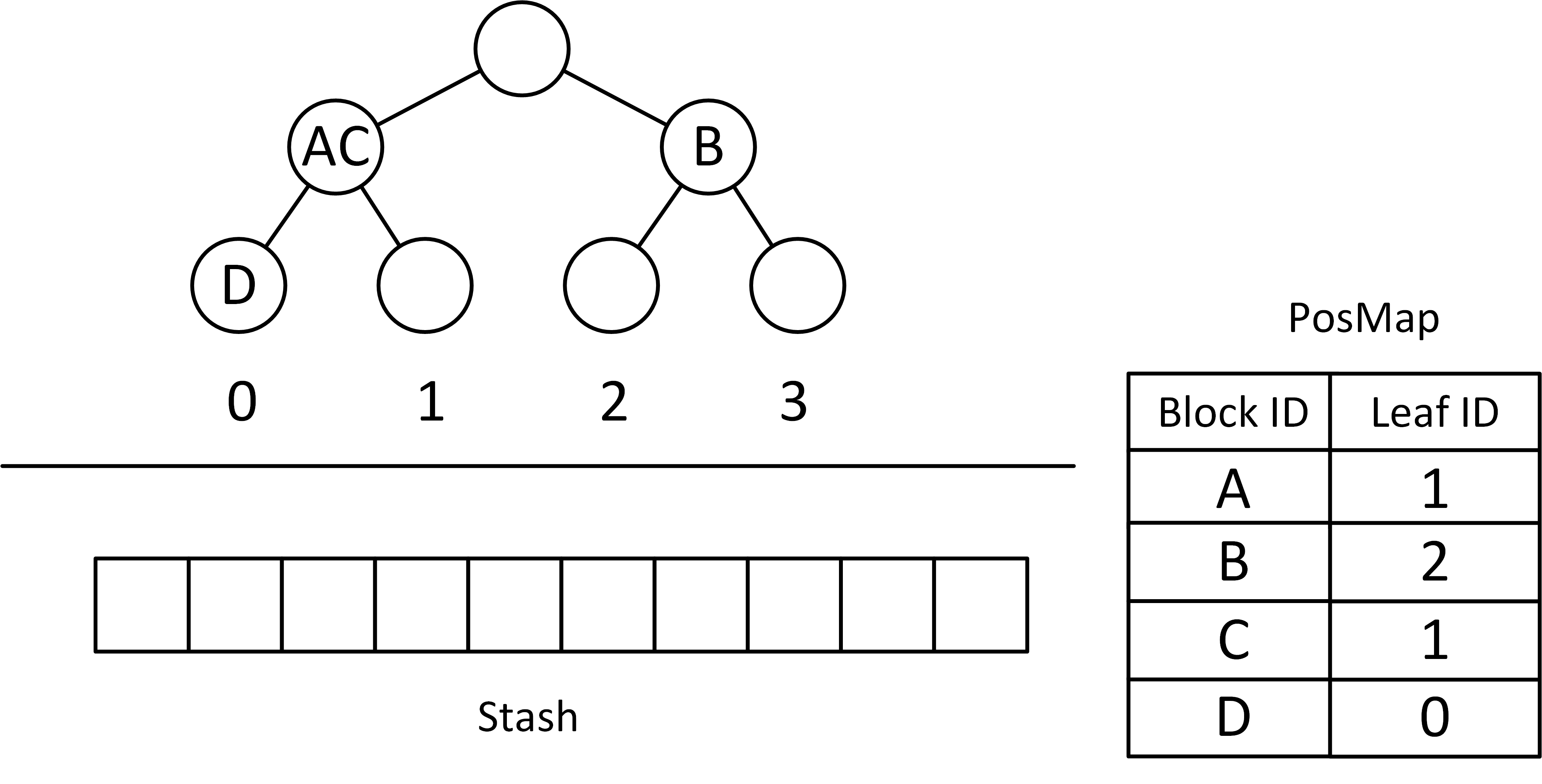}\label{fig:path_oram_example_after_write_back}}%
        \caption{\textbf{Illustration of Path ORAM accesses and important hardware structures involved.}}
        \label{fig:path_oram_example}
\end{figure*}

In this section, we will introduce necessary background to understand this work. To that end, first, we discuss background on non-volatile memories (NVMs) and the endurance limit they manifest followed by a discussion of Start-Gap~\cite{Start-Gap}, a state-of-the-art wear-leveling mechanism. Then, we discuss the threat model we assume in this work. After that, we provide background on the most state-of-art implementation of ORAM primitive, Path-ORAM~\cite{path_oram}. Finally, we discuss how the Path-ORAM algorithm stresses the endurance of  NVMs.

\subsection{Non-Volatile Memories}
\label{sec:background_motivation_pm}

Emerging  {non-volatile} memories  {(NVMs)}, including PCM, STTRAM and ReRAM \cite{PCM, STTRAM, ReRAM} can provide several desirable properties like high density, non-volatility and an access latency close to DRAM~\cite{pm_comparable_performance_with_dram}. Further, as  NVMs are byte-addressable (they provide load/store interface\cite{PM_load_store}) and non-volatile, they can make it possible to host recoverable data structures in  NVMs instead of accessing them from disk at steeper performance and energy costs. 

While  NVMs have several desirable characteristics making their inclusion in future memory systems possible, several challenges remain. One of the challenges is about their limited write endurance. A typical write endurance ranges from $10^7 $ to $10^8$ \cite{pcm_lifetime} due to  physical properties of  NVM memory cells. Once the limit is passed, data errors may raise because memory cells may no longer be able to change their states, which can in turn lead to system failure if sufficient memory lines reach endurance limit. This happens when the number of spare memory lines is lower than memory lines that have incurred more writes than dictated by the endurance limit. 



Several prior works have been proposed to solve the limited write endurance problem of  NVMs. These approaches can typically be categorized into write reduction techniques and wear-leveling techniques. Write reduction techniques focuses on reducing the number of  NVM writes. For example, Qureshi et al. \cite{PCM_as_second_storage} proposed using PCM as an extended storage of DRAM. Gogte et al. \cite{one_percent_lifetime} proposed using software management system to migrate frequently accessed pages from  NVM to DRAM. On the other hand, wear-leveling techniques focus on spreading memory writes uniformly across all  NVM memory space. We discuss several prior wear-leveling techniques and compare them to our proposed work in Section~\ref{sec:related}.  

\subsection{Baseline Start-Gap}
Start-Gap divides  NVMs into equal-sized groups. Each group reserves an empty cache line. The CPU maintains a start register and a gap register for each group. Each memory address will be translated based on the value of start and gap registers. The CPU maintains a counter for each group to store the number of writes to that group so far. Upon the counter reaching a threshold, the gap register will be updated, so one of the cache lines in that group will be mapped to a new address. Once the gap register finishes looping through the entire group, the start register will be updated. The algorithm also employs an address randomizer which maps logical addresses to intermediate addresses before further translating them to physical addresses with start and gap registers.

\subsection{Threat Model}
\label{sec:background_motivation_threat_model}

    
We assume a secure processor where applications can harness isolated execution~\cite{secure_processor_1}. We assume that the execution of an application along with its data in the processor structures (registers, caches, on-chip networks) are secure and isolated from other applications. We assume a system where  NVM is employed as the main memory. We assume the  NVM is untrusted and to defend against cold (re)boot attacks~\cite{cold_boot_attack}, data is encrypted in memory. Further, we also assume an adversary with physical access to the computers and as such, capable of probing the off-chip memory bus~\cite{bus-snooping} to learn the address trace and data being communicated between the processor and  NVM. Prior works~\cite{msr_access_trace_info_leak} have shown that even with encrypted data, an attacker can ascertain sensitive secrets about an application simply by observing the address trace of the application. We assume prior solutions for mitigating page-fault side-channel~\cite{invisipage, sanctum}. Finally, prior works which address other important vulnerabilities such as power~\cite{side_channel_power}, thermal~\cite{side_channel_thermal}, program execution time~\cite{side_channel_execution_time} side-channels, and leaks via communication patterns over the network~\cite{side_channel_internet_1,side_channel_internet_2} do exist and we assume addressing these vulnerabilities to be outside the scope of this paper. 

\subsection{Path Oblivious RAM}
\label{sec:background_motivation_path_oram}

Oblivious RAM \cite{oram_1}, is a cryptographic primitive which makes a memory access trace computationally indistinguishable from a random access trace of the same length. While several realizations of this primitive have been proposed to date~\cite{oram_previous_work_1, oram_previous_work_3, oram_previous_work_3, oram_previous_work_4}, a backbone for them is Path ORAM~\cite{path_oram}, the most practical implementation of ORAM primitive. In this section,  we provide relevant background on Path-ORAM algorithm (henceforth simply referred to as ORAM).

ORAM arranges memory into a balanced binary tree structure, and each node in the ORAM tree contains a fixed number of slots where each slot can store a single data block (typically a single cache block). ORAM tree has associated with it a \textit{utilization} factor, referring to the fraction of actual data blocks over total blocks in the tree. Non-data blocks are filled with dummy blocks. Each real data block has a corresponding leaf ID, which indicates that this data block is stored on the path from the root node to the leaf corresponding to the leaf ID. This mapping information is maintained by a \textit{position-map} (PosMap) structure. All blocks, regardless of real data block or dummy block, are encrypted before being stored into memory. Encryption metadata, such as block ID (the slot within each node), leaf ID, as well as encryption counter, are also maintained by the ORAM tree. ORAM also uses an on-chip structure, \textit{stash}, to aid in its working. 

Next, we use a 3-level ORAM tree in Fig. \ref{fig:path_oram_example} as an example to show the workings of ORAM. Consider the scenario when the application issues a read request to data block C. Fig \ref{fig:path_oram_example_init} shows the initial status of the ORAM tree, stash and PosMap. On receiving the read request, the algorithm first looks up PosMap to find the leaf ID that the block C corresponds to, which is 0 (leaf nodes in the tree are labeled from 0 to 3). Per the ORAM primitive, each program memory access is translated to a read and write of a path from root node to the leaf associated with the block. As such, the algorithm proceeds to read and write path to leaf 0 as depicted in Fig. ~\ref{fig:path_oram_example_init}. To do so, ORAM issues read operations to all nodes along the path from the root node to leaf 0 and loads real data blocks (C and D), into the stash. Post the path read the needed data block is in the stash and is decrypted and supplied to the processor. At this moment, ORAM will remap block C to a newly generated random leaf ID, which is 1 in this case, as shown in Fig. \ref{fig:path_oram_example_after_read}. This remapping confirms that the next access to block C accesses a completely random path in the ORAM tree. Finally, ORAM writes back the same path it read from, draining re-encrypted blocks from the stash as deep as possible. In this example, block D is mapped to leaf 0, which is fine to be pushed into leaf 0. Block A and C are mapped to leaf 1, so the deepest node they can be stored in is the parent node of leaf 0 and 1.

By re-mapping the data block each time it is accessed to a random leaf, ORAM allows for a completely new set of blocks are accessed each time the same data block is accessed. This, along with the fact that each access causes re-encryption of all accessed blocks helps hide address trace of the application.

\subsection{Challenges in realizing ORAM for NVMs}
\label{sec:background_motivation_pm_oram_challenges}

\begin{figure}[t]
    \centering
    \includegraphics[width=0.6\linewidth]{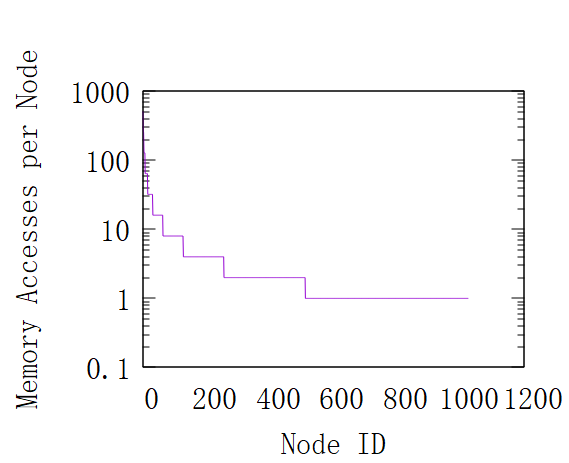}
    \caption{\textbf{ORAM's exponential write distribution:} The x-axis shows the node ID. The y-axis shows the corresponding expected number of writes to each node in log scale normalized to the expected number of writes to a leaf node. We assume that the tree has 10 levels and that nodes closer to root are assigned with smaller IDs. }
    \label{fig:path_oram_write_distribution}
\end{figure}

We discuss in this section how, by its very nature, the ORAM primitive severely stresses endurance of  NVMs. This is so, as ORAM manifests a skewed write access pattern as shown in Figure~\ref{fig:path_oram_write_distribution}. 
As discussed in section~\ref{sec:background_motivation_path_oram}, memory is organized as a binary tree with ORAM and memory accesses are transformed into read/writes of paths in the tree. This translates to the scenario as depicted in Figure~\ref{fig:path_oram_write_distribution}, where the nodes closer to the root (lower Node-ID) are written far more often than nodes closer to leaf nodes (higher Node-ID). 
From the root node, its two child nodes are accessed with equal probability due to the nature of ORAM, so each child node is accessed with half frequency of the root node. This access property applies to all nodes, which causes the number of writes per node in level to decrease exponentially as the level increases linearly. Therefore, we see that leaf nodes are written far less frequently than the root node, while occupying half of the tree. 

\begin{figure}[t]
    \centering
    \includegraphics[width=0.6\linewidth]{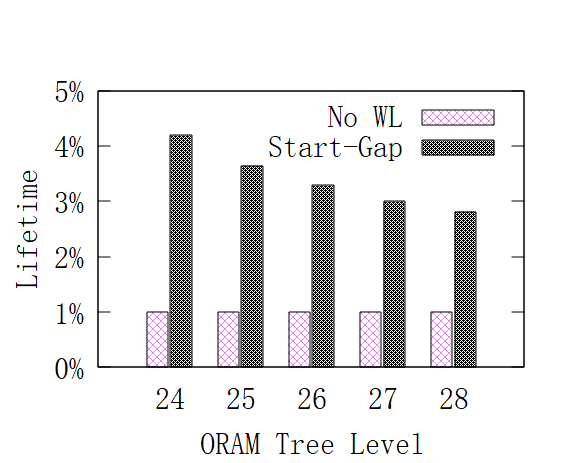}
    \caption{\textbf{ORAM significantly reduces  NVM lifetime:} This figure shows the lifetime of an ORAM enabled  NVM device with and without Start-Gap~\cite{Start-Gap}. We assume a region-based Start-Gap with randomized grouping and 256 groups as that performs the best wear-leveling in our sensitivity analysis.}
    \label{fig:path_oram_lifetime_baseline}
\end{figure}

We depict the stress ORAM's write distribution causes on  NVM endurance in Fig.~\ref{fig:path_oram_lifetime_baseline} (please see Section~\ref{sec:evaluation} for details on our methodology). In Fig.~\ref{fig:path_oram_lifetime_baseline} we depict system \textit{lifetime} for a  NVM that implements ORAM. We define system lifetime to be: 

\begin{equation*}
    Lifetime = \frac{\mbox{ \#memory writes $\times$ ORAM tree path length}}{\mbox{\#cache lines $\times$ Wmax}}
\end{equation*}

As every program memory access in ORAM translates to read and write of a path in ORAM tree (as explained in Section~\ref{sec:background_motivation_path_oram}), in the above equation, the numerator calculates the total reads/writes to  NVM which implements ORAM before system failure. The denominator calculates the ideal number of reads/writes possible to memory with perfect wear-levelling, where Wmax refers to the endurance of a cache line, e.g., the number of writes to a cache line before it fails. We chose Wmax to be $10^7$. We consider a  NVM failure and consequent system failure when more than 1\% of cache lines have reached endurance limit, similar to a mechanism used in \cite{one_percent_lifetime}. 
Further, we show both the lifetime without wear-leveling and in presence of state-of-art wear-levelling solution, \textit{Start-Gap}~\cite{Start-Gap}, As Fig.~\ref{fig:path_oram_lifetime_baseline} depicts, both configurations achieve less than 4\% lifetime and furthermore, lifetime drops with larger ORAM tree size (= larger  NVM memory size). Emerging  NVMs are deployed in terabytes \cite{terabytes_PM}, which would translate to a very large tree, e.g., 32-level tree. As a result, it is expected that the lifetime will be poor. 
This makes it impractical to implement ORAM for  NVMs and consequently, impractical to address memory bus side channel for  NVMs.

\begin{figure*}
    \centering
    \includegraphics[width=.8\linewidth]{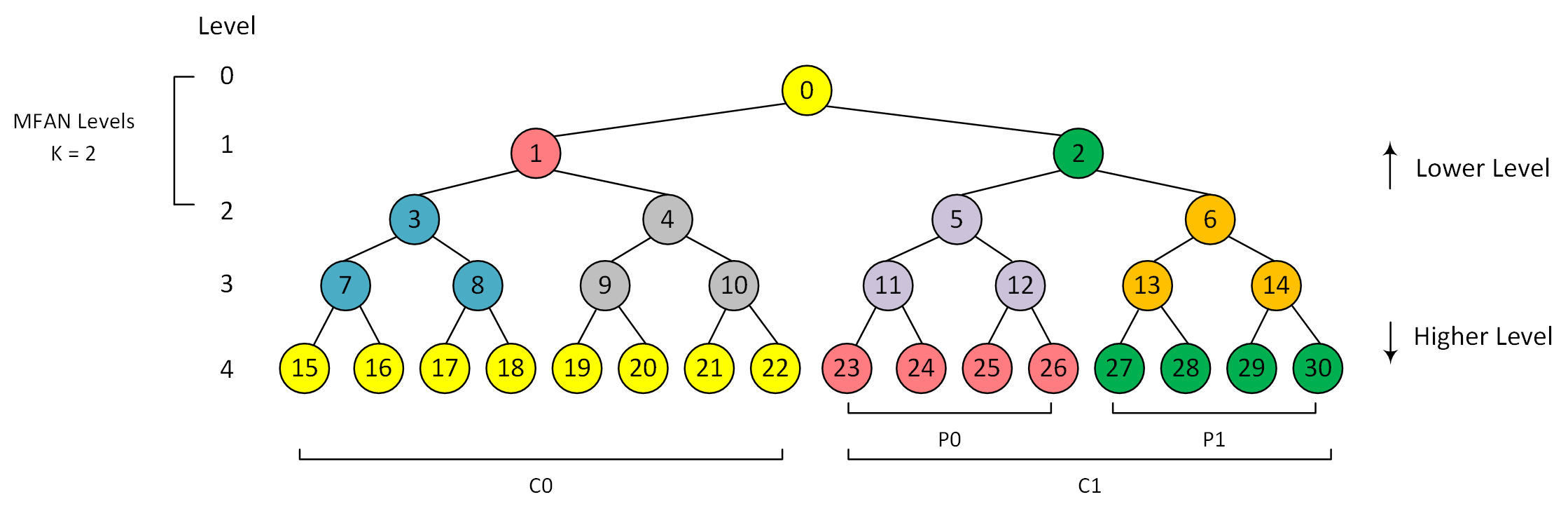}
    \caption{\textbf{Example static group partitioning with \DESIGN:} A ORAM tree with 5 levels and 31 nodes partitioned into 7 groups, each identified with a different color. MFAN level,  {$K$} (defined in section \ref{sec:grouppartitioning}), is 2. }
    \label{fig:partition_example}
\end{figure*}

\section{Egalitarian ORAM - Insight}
\label{section:key_idea}
In this section, we propose the inspiration for \DESIGN, our proposed wear-levelling algorithm to tackle endurance pressure of ORAM primitive for  NVMs. 

Fig. \ref{fig:partition_example} shows a simple 5-level ORAM tree with 31 nodes. In this example, we label each node with a unique node ID. Nodes closer to the root node are assigned smaller node IDs.
$N$ ORAM accesses to this tree result in a total of $5*N$ node reads and writes, $N$ for each of the 5 levels.
However, since levels 0 to 4 contain 1, 2, 4, 8, 16 nodes respectively, the number of reads and writes to each of the nodes in these levels  are $N$, $\frac{N}{2}$, $\frac{N}{4}$, $\frac{N}{8}$, $\frac{N}{16}$ respectively. The goal of an ideal wear-leveling algorithm would be to confirm that each of the 31 physical node locations see the same number of writes, i.e., $\frac{5*N}{31}$. This goal can be achieved by periodically remapping logical nodes to different physical nodes so that the writes are as close to uniformly distributed as possible. To minimize performance overheads from the re-mappings, we have to ensure that there are (i) only as many re-mappings as necessary and (ii) the indirection costs associated with routing accesses to logical nodes to the appropriate physical nodes must be kept low.

\DESIGN is based on the observation that nodes in levels 0, 1, and 2 are accessed more frequently than the ideal and the remaining nodes in levels 3 and 4 are accessed less frequently.
If we were to form a group with the root node from level 0 and eight nodes from level 4 and assume ideal wear-leveling within the group, we can achieve average number of writes per node of $\frac{N}{6}$, coming within 3\% of the ideal ($\frac{5*N}{31}$). Similarly, a group formed with one node from level 1 and four nodes from level 4 comes within 7\% of the ideal. As we discuss in detail later, continuing this process of grouping one frequently written node (referred to as the Most Frequently Accessed Node, MFAN, henceforth) from a level close to the root with some partner nodes (referred to as the Partner Nodes, PNs, henceforth) from a relatively infrequently written level and wear-leveling within each of the groups can get us very close to ideal wear-leveling.  {We define such in-group wear-leveling as MFAN movement.} 

Note that prior state-of-art wear-levelling solution, Start-Gap \cite{Start-Gap}, also proposed breaking down memory space into groups and performing wear-levelling in each group. However, we augment this basic idea in several important ways. When handling ORAM's exponential write distribution, Start-Gap falls short due to its design choice of breaking down the memory addresses into randomized, fixed-sized, wear-leveling groups. With fixed sized groups, if a group happens to contain a very heavily written ORAM node (say the root node), then that group is likely to wear out much sooner than other groups that contain only relatively infrequently written nodes (say only the leaf nodes). This imbalance limits the lifetime achieved with Start-Gap on ORAM-style access patterns. However, one of the advantages of ORAM-style access pattern is that given its randomization and the binary-tree arrangement, the fraction of total writes incurred by each node is known apriori just by knowing the size of the ORAM tree. With this knowledge, it is possible to statically partition the ORAM tree into groups such that average number of writes per node incurred across all the groups is similar. Our central idea is to form variable sized groups instead of fixed-size groups in an ORAM primitive aware manner (unlike randomized in Start-Gap). This design choice allows very frequently written nodes like the root node to be grouped with a large number of infrequently nodes and reduce the average number of writes per node. However, having variable sized groups creates challenges in keeping track of which nodes belong to which groups and the wear-leveling movements within each group. We discuss our solutions to these challenges next.

\section{Egalitarian ORAM}
\label{sec:algorithms}

In this section, we discuss two parts \DESIGN is composed of: 1) the partition algorithm to group ORAM tree nodes and 2) the node remapping algorithm, which efficiently remaps nodes within the same group.

\subsection{Static group partitioning}
\label{sec:grouppartitioning}

The first part of \DESIGN involves a static partition of the ORAM tree. We partition the ORAM tree into many groups with the following goals: (i) the average number of writes per node in each group approaches the average number of writes per node in the entire tree and (ii) within each group, there exists exactly one node being written more frequently than all other nodes (MFAN as defined in section \ref{section:key_idea}), and all other nodes are expected to be accessed equally (partner nodes, as defined in section \ref{section:key_idea}). This goal is to simplify the hardware implementation of \DESIGN as we will explain in \S\ref{sec:MFAN_movement_algorithm}.

Given the two principles, we can design an algorithm to partition our ORAM tree. For an ORAM tree with $L$ levels (with level 0 being the root node), we will choose a threshold level, $t$, such that all nodes belonging to levels $0$ to $(t-1)$ are grouped with level $(L-1)$, which is the leaf level. Exactly how we choose the value of $t$ will be described later in this section. Then, nodes in the leaf level are further divided into $t$ equal-size chunks, and assign them with chunk IDs, defined as $C_i$, where $0 \leq i \leq t$. Nodes in $C_i$ are paired with nodes in level $i$. Use Fig. \ref{fig:partition_example} as an example, we will choose $t = floor(5/2) = 2$ as the first step, and then divide nodes in level 4 into 2 chunks, $C_0$ and $C_1$.
$C_0$ nodes (yellow nodes in level 4) are grouped with the root node (also yellow), while the two nodes in level 1 (nodes 1 and 2, red and green respectively) are grouped with half of the nodes in $C_1$ each (red and green respectively). We denote each subset nodes in $C_1$ as Parts and assign them with Part IDs.
So far, we have formed groups for nodes in level 0 and 1 and level 4, that leaves the nodes in levels 2 and 3 to be grouped.
As can be seen from Fig.~\ref{fig:partition_example}, the nodes in levels 2 and 3 can be viewed as four separate, smaller ORAM trees (in blue, grey, purple, and orange colors).
Our partitioning strategy can be recursively applied to these intermediate, smaller ORAM trees until the entire larger tree has been partitioned.
One important characteristic in the groups formed by our algorithm is that each group has one MFAN (most frequently accessed node) and the rest of the partner nodes all belong to the same, less frequently accessed level. With this characteristic, we define   {$K$} as the highest level that contains MFAN nodes, called MFAN level. In this example,   {$K=2$}. This value is crucial when discussing how to trigger MFAN movement in section \ref{sec:triger_mfan_movement} and performance overhead in section \ref{sec:performance_overhead}.

{The last aspect that we have not yet discussed is how to choose an appropriate threshold level $t$. 
Assuming perfect intra-group wear-leveling, we experimented with different values of $t$. 
We found that threshold level value of $\frac{L}{2}$ gives us the lowest number of writes per node in the most vulnerable group, making it the optimal threshold level value in our algorithm.
}


\subsection{Node Remapping}
\label{sec:node_remapping}
While the groups are decided statically, at runtime, the wear-leveling algorithm has to perform the following actions: (i) decide when to perform the MFAN movements for each of the groups and (ii) accurately maintain the mapping of logical node IDs and corresponding physical locations accounting for all the MFAN movements performed.
Apart from being functionally correct, we also aim to perform the above two actions with the high performance and low storage overheads. Next, we describe how we perform both actions.

\subsubsection{MFAN Movement Algorithm}
\label{sec:MFAN_movement_algorithm}
In this section, we will discuss how MFAN movements are handled. 
Notice for a group $g$, when an MFAN movement occurs, we need to perform a data swap between MFAN and one of its partner nodes. We will use Fig. \ref{fig:partition_example} as an example to illustrate the MFAN movement algorithm.

\begin{figure*} 
     \subfloat[Initial Status of the red group.]{\includegraphics[width=0.6\columnwidth]{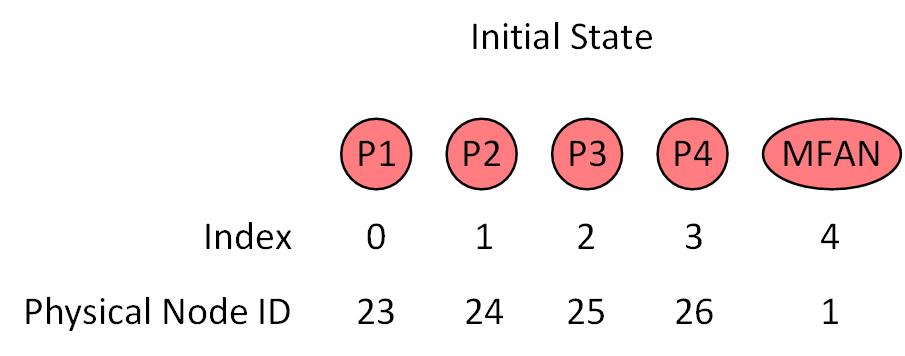}\label{fig:MFAN_movement_init}}%
    \hspace{0.5cm}
     \subfloat[Status of the red group after swapping MFAN with P4.]{\includegraphics[width=0.6\columnwidth]{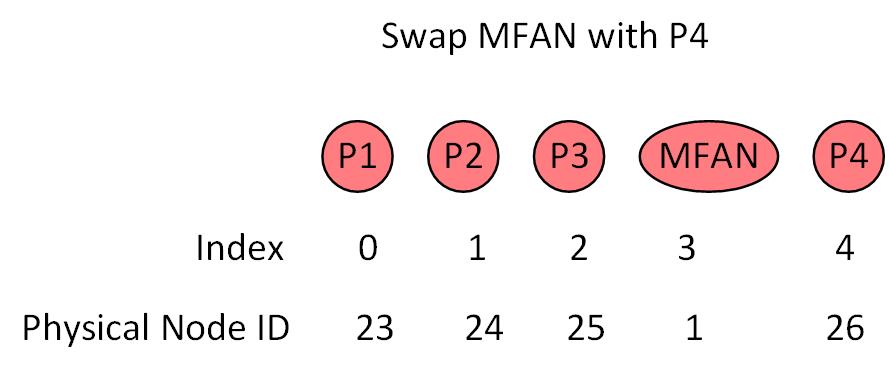}\label{fig:MFAN_movement_1}}%
    \hspace{0.5cm}
     \subfloat[Status of the red group after swapping MFAN with P3. ]{\includegraphics[width=0.6\columnwidth]{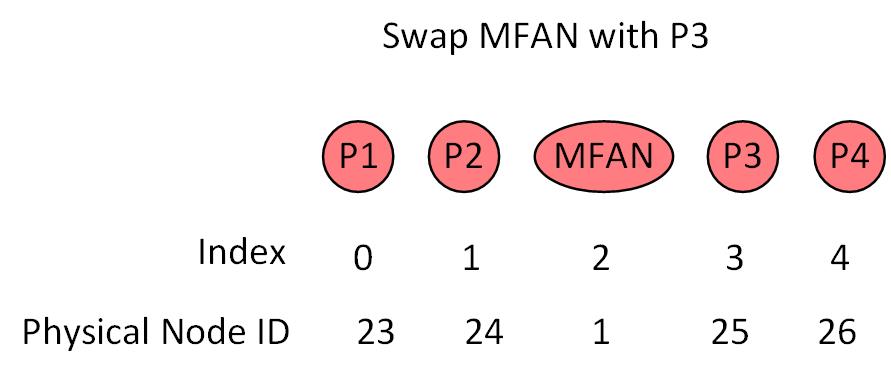}\label{fig:MFAN_movement_2}}%
        \caption{\textbf{Illustration of MFAN movement in a group with one MFAN and four partner nodes.}}
        \label{fig:MFAN_movement_example}
\end{figure*}

Consider the red group, it has fives nodes in total: the MFAN (node 1 in level 1) and four partner nodes (P1, P2, P3, P4, nodes 23, 24, 25, 26 respectively in level 4). 
For illustration purposes, we can visualize these five nodes arranged in an array in the order P1, P2, P3, P4, MFAN, as shown in Fig. \ref{fig:MFAN_movement_init}.
After a threshold number of writes are performed to this group, we swap the MFAN with the node on its left in the array (wrapping around when necessary).
Fig. \ref{fig:MFAN_movement_1} shows the state of the array after swapping P4 and MFAN.
Before the swap, all the writes to MFAN were handled by the physical node 1 and all the writes to P4 were handled by the physical node 26.
After the swap, all the writes to MFAN will be handled by the physical node 26 and vice-versa.
So, this swap effectively wear-levels between physical node 1 and 26.
Similarly, once we hit the threshold number of writes in the new arrangement, the MFAN, now at physical node 26, will be swapped with the partner node to its left (P3), now at physical node 25, as shown in Fig. \ref{fig:MFAN_movement_2}.
When we are ready to make the next swap, physical nodes 25, 26, and 1 all would have seen about the same number of writes, in other words they have been wear-levelled.
We continue this MFAN movement within the group to provide effective intra-group wear-levelling.

A few important things to notice about this intra-group wear-levelling approach:\\
\textbf{Simple and ideal intra-group wear-levelling:} Given the way in which we construct our groups, each group has one MFAN and many partner nodes where each partner node is expected to incur the same number of writes. 
The partner nodes all being equivalent (from a write frequency point-of-view) allows us to achieve ideal intra-group wear-levelling by simply moving the MFAN, no other nodes need to be moved unless they are being swapped with the MFAN.\\
\textbf{Deterministic locations:} One requirement for swap-based wear-leveling approaches is to be able to accurately locate the nodes after a series of swaps. 
One approach to tracking the physical locations of logical nodes is to maintain a mapping table for the indirection.
However, given that we move the MFAN in a deterministic pattern and at fixed intervals (every threshold number of writes), just tracking the number of writes to the group is enough to exactly determine the physical location to which the MFAN is currently mapped to.
Furthermore, since the MFAN is the only node that gets swapped, the location of the partner nodes can also be formulaic-ly determined simply based on the number of number of writes incurred by the group.
Specifically, the following two formulae will allow us to accurately determine the exact physical locations of the MFAN and each of the partner nodes based on the number of writes to the group.

For the new array index of MFAN, we can compute it with the following equation:
\begin{equation}
    MFAN = (N_g - 1) - s_g \% N_g
    \label{eq:MFAN_computation}
\end{equation}
where $N_g$ is the size of this group, and $s_g$ refers to the number of swaps happened in this group. 
 

For any partner node whose initial array index is $y$, its new array index $y'$ can be computed with
\begin{equation}
     y' = \left(y + floor(\frac{s_g + y}{N_g-1})\right) \% N_g \mbox{ (y $\neq N_g-1$ )}
     \label{eq:partner_node_computation}
\end{equation}


Notice that for both equation \ref{eq:MFAN_computation} and \ref{eq:partner_node_computation}, the input and output value are both array indices. For example, if we want to compute the location of MFAN shown in Fig. \ref{fig:MFAN_movement_init}, we can achieve this by feeding $N_g = 5$ and $s_g = 0$ into equation \ref{eq:MFAN_computation}, which gives us 4. This value 4 needs to be mapped to physical node 1, because it is the last element of the array, which is mapped to node 1. Similarly, after 1 swap operation, equation \ref{eq:MFAN_computation} will tell us that MFAN locates in array index 3. Since element 3 is not the last element in the array, we know that this value must be mapped to one of the partner nodes in the group. We can find its physical location by adding 3 with a $OFFSET$ value, where $OFFSET$ denotes the the node ID that array index 0 being mapped to. In this example, $OFFSET = 23$, so MFAN now locates in physical node 26. 

\begin{figure}[t]
    \centering
    \includegraphics[width=.8\linewidth]{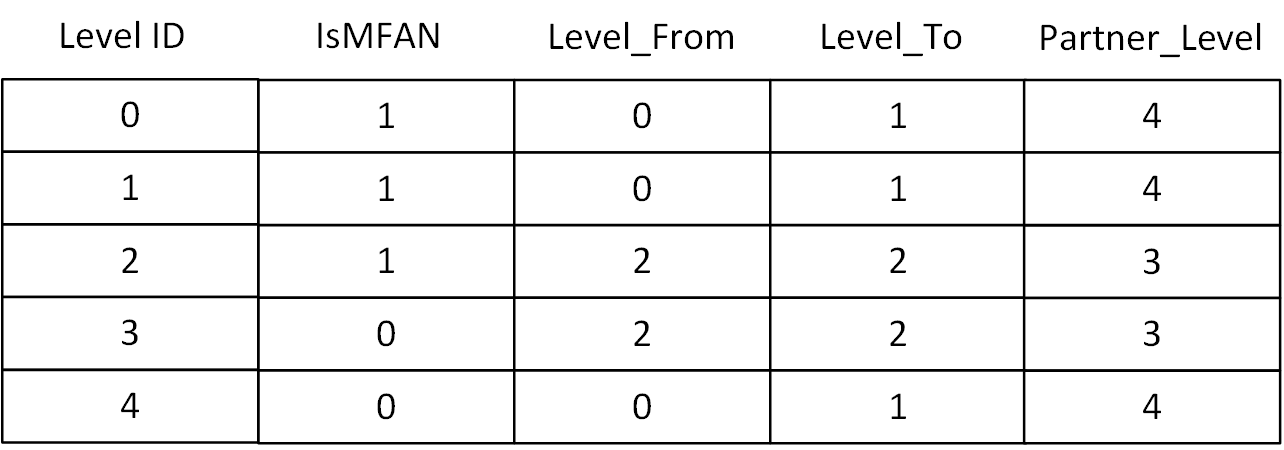}  
    \caption{\textbf{A lookup table illustration:} A sample lookup table stored on chip, which is used to compute $OFFSET$.}
    \label{fig:on_chip_look_up_table}
\end{figure}

The $OFFSET$ value varies among groups. It is unacceptable if we store all $OFFSET$ values on chip, considering the huge number of groups we have. Fortunately, it is not hard to show that the value of $OFFSET$ for an arbitrary group can be computed by the following information: 1) the level that all partner nodes reside in, denoted as $l_p$, which is 4 in this example; 2) the MFAN ID, specified by ($i$, $j$), which refers to the node in $i^{th}$ level, $j^{th}$ node in that level; 3) the number of levels sharing the same partner level, denoted as $l_s$. We propose storing such information in an on-chip lookup table, as shown in Fig. \ref{fig:on_chip_look_up_table}. In Fig. \ref{fig:on_chip_look_up_table}, column 'IsMFAN' contains a single bit indicating whether nodes in a level are MFANs in their groups. Column 'PartnerLevel' denotes the level containing partners nodes, which is $l_p$. Column 'LevelFrom' refers to the starting level being paired with $l_p$, and column 'LevelTo' is the ending level being paired with $l_p$. Essentially, variable $ l_s$ can be computed by subtracting 'LevelTo' by 'LevelFrom' for a given level ID. Then, $OFFSET$ can be easily computed by

\begin{dmath}
    OFFSET=   2^{l_p}-1 + ChunkID\times ChunkSize 
              + PartID \times PartSize
    \label{eq:offset_computation}
\end{dmath}

where term $2^{l_p}-1$ is the first node in level $l_p$; ChunkID can be computed by value $i$, which is 1 in our example; ChunkSize can be found by the value of $l_s$ by dividing the number of nodes in level $l_p$ by $l_s$, which is 8 in our example; PartID can be found with the value of $j$, which is 1 in our example; and finally, PartSize can be found by dividing ChunkSize by the number of nodes in level $i$, which is 4 in this example. Plug all values into equation \ref{eq:offset_computation}, we get $OFFSET=23$.  

This deterministic movement pattern implies that we do not need an expensive mapping table for each group, instead we simply can implement the circuitry necessary to calculate the exact locations of each of the nodes in the ORAM controller.

\subsubsection{Triggering MFAN Movement}
\label{sec:triger_mfan_movement}
To trigger gap movement, Start-Gap \cite{Start-Gap} reserves an on-chip counter for each group in the  NVM. Each counter stores the number of writes to its corresponding group. When a counter reaches a threshold, $X$, Start-Gap will trigger gap movement on that group. Start-Gap defines $X$ as wear-leveling frequency \cite{Start-Gap}. We will use the same definition here. In our case, we employ a similar mechanism in our design. Ideally, we could store on-chip counters for all groups, but this is unrealistic because of the extremely large number of groups generated by our partition algorithm. Fortunately, we observe strong correlations among the number of writes to groups. Recall that ORAM assumes a random leaf node access. This random number is assumed to be uniformly distributed. Therefore, we can compute the expected number of writes to each node, and therefore, to each group given the static ORAM tree partitions. Then, if we reserve one global counter which stores the total number of ORAM accesses so far, then we could compute the expected number of memory writes to any group. In addition, we found that for a group whose MFAN is located in level $i$, denoted as $g(i)$ ($g(i)$ may refer to $2^i$ groups), the expected number of writes to group $g(i+1)$ is around half of $g(i)$. This is true because the MFAN node in $g(i+1)$ receives half number of writes compared to the MFAN node in $g(i)$, and the number of partner nodes in $g(i+1)$ is also half of the number of partner nodes in $g(i)$. If partner nodes in $g(i)$ and those in $g(i+1)$ are located in the same level, then the expected number of writes to group $g(i+1)$ is exactly half of $g(i)$. 

Since the expected number of writes to a group is determined by its MFAN, then we could simply reserve 1 on-chip counter, $Ctr$, to store the total number of writes to the  NVM. For $g(0)$, we perform MFAN movement with a period of $X$, and for $g(1)$, we perform MFAN movement with half frequency compared to $g(0)$, etc. The equivalent expression is that for every $X$ ORAM accesses, all $g(0)$ will perform MFAN movements; for every $2X$ ORAM accesses, all $g(0)$ and $g(1)$ will perform MFAN movements, etc. Therefore, the MFAN movement triggering algorithm can be formulated as shown in algorithm \ref{algo:WL_algorithm}:

\begin{algorithm}
\SetAlgoLined

\KwData{ Counter  $Ctr$, MFAN Level   {$K$}, Wear-Leveling Frequency $X$}
    \KwResult{A set, $S$, which contains level IDs. All groups whose MFANs locates in these levels needs to perform MFAN movements}
    
    On an ORAM request:                 \\
    \If {$Ctr \mbox{   \%   } X = 0 $}{
        $k \gets 0$              \;
        
        \While {$k < K \And Ctr \mbox{  \%   } 2^{k+1} X \neq 0$} {
            $k \gets k+1$                             \;
        } 
    }
    $S \gets \{0, 1, ..., k\}$            \;
    $Ctr \gets Ctr+1$                               \;
    return ($S$, $Ctr$)                   \;

 \caption{MFAN Movement Triggering Algorithm.}
  \label{algo:WL_algorithm}

\end{algorithm}

In algorithm \ref{algo:WL_algorithm}, $Ctr$ refers to the on-chip counter;   {$K$} refers to the MFAN level. In Fig. \ref{fig:partition_example}, the value of   {$K$} will be 2. The return value, $S$, essentially states that MFAN movements need to be performed on $g(0)$, $g(1)$ ... $g(k)$. 

\begin{figure}[t]
    \centering
    \includegraphics[width=.7\linewidth]{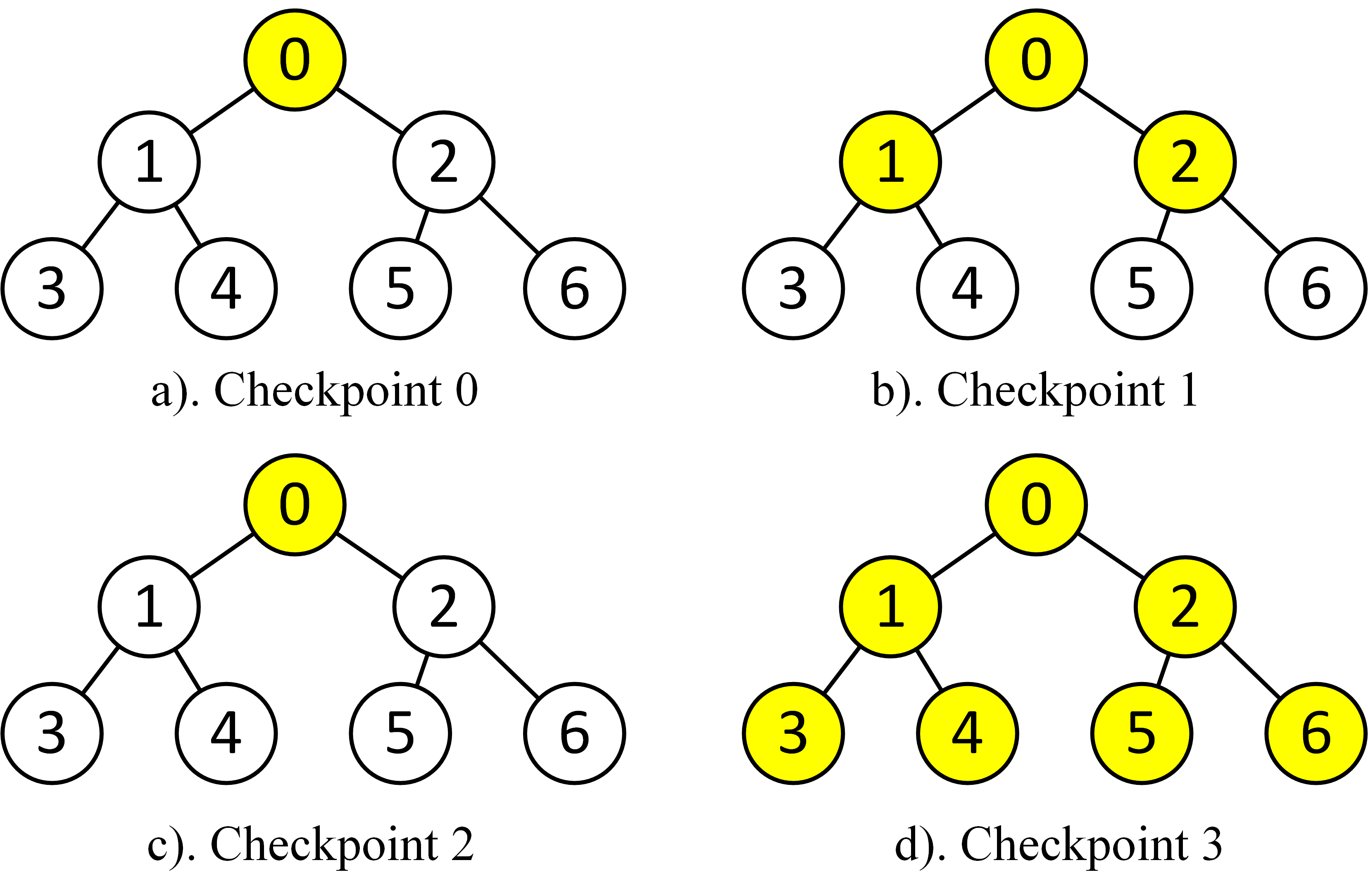}
    \caption{\textbf{Illustration of the naive MFAN movement triggering algorithm:} For different checkpoints, there is a significant imbalance in the number of groups that need to perform MFAN movements.}
    \label{fig:wl_triggering_algorithm_original}
\end{figure}

\begin{figure}
    \centering
    \includegraphics[width=.7\linewidth]{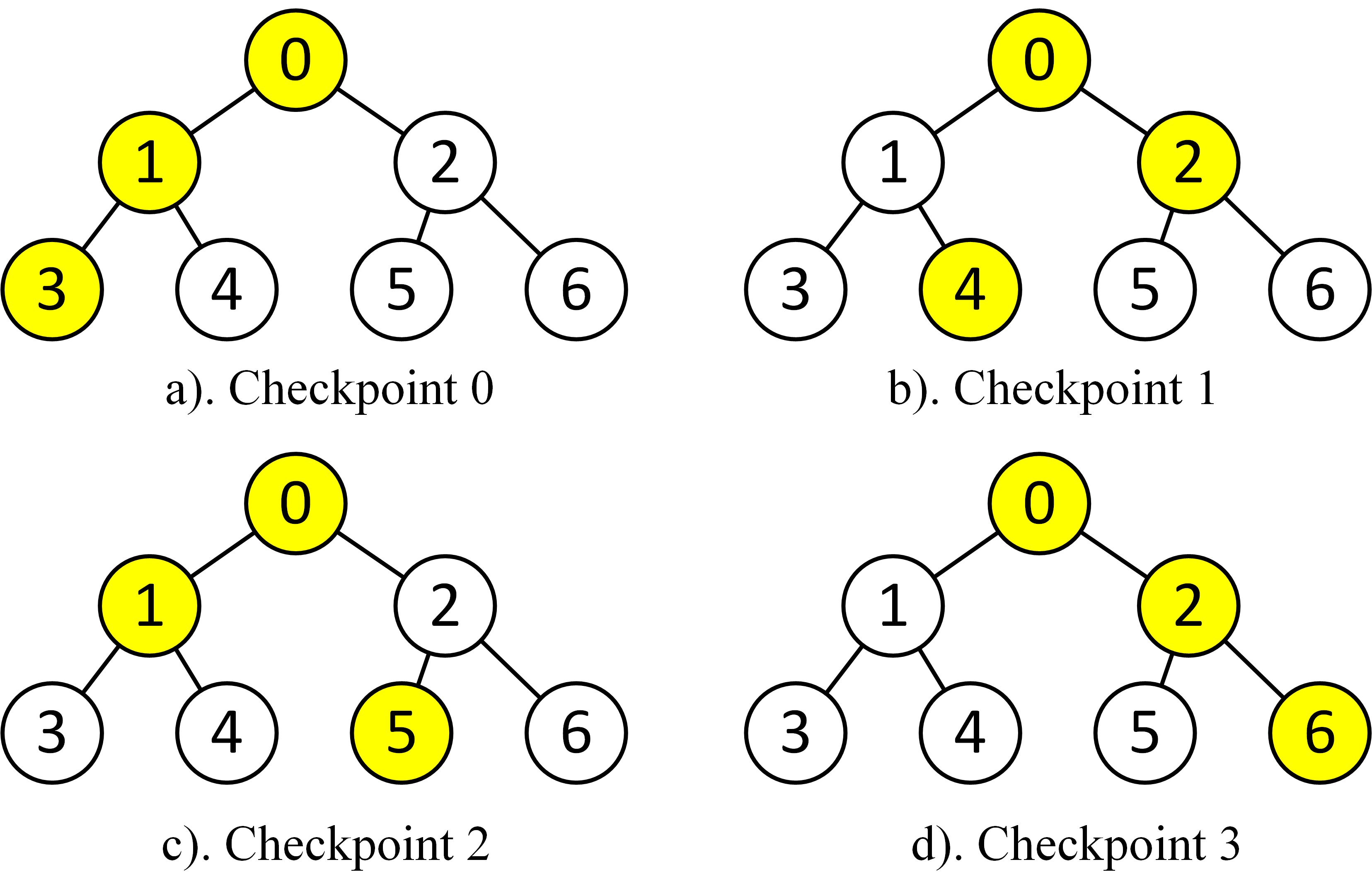}
    \caption{\textbf{Illustration of a balanced MFAN movement triggering algorithm:} In the new algorithm, exactly 3 groups need to perform MFAN movements at each checkpoint.}
    \label{fig:wl_triggering_algorithm_optimzied}
\end{figure}

One of the drawbacks of algorithm \ref{algo:WL_algorithm} is that when the counter hits $2^K X$, then all groups need to perform MFAN movements.
The large number of MFAN movements performed at the same time could starve out actual ORAM accesses that need to be performed.
For example, there are $2^{28}-1$ groups for a 32-level ORAM tree.
To avoid this concern, we improve our MFAN movement triggering algorithm to stagger the MFAN movements.
We shall work through an example to illustrate the problem and our solution. For simplicity, we define a checkpoint as each time $Ctr \% X = 0$ satisfies. The first time $Ctr$ reaches $X$, we define it as checkpoint 0. Fig. \ref{fig:wl_triggering_algorithm_original} shows such an example. Suppose we have an ORAM tree with MFAN level   {$K=2$}. Then, at checkpoint 0, only the group whose MFAN node is the root node needs to perform MFAN movement. At checkpoint 1, 3 groups colored in yellow needs to perform MFAN movement. At checkpoint 2, only the root group needs to perform MFAN movement. This is because $Ctr = 3X$, so $k$ remains 0. At checkpoint 3, since $Ctr = 4X$, so $k$ is 2 now, we need to perform MFAN movement for all 7 yellow groups. This imbalance of the number of MFAN movements at each checkpoint could cause an ORAM request to be blocked for a long time. Alternatively, we observe that we can solve this problem simply by rearranging the order of doing these MFAN movements, as shown in Fig. \ref{fig:wl_triggering_algorithm_optimzied}. We notice that we can perform exactly three MFAN movements at each checkpoint. At checkpoint 0, we perform MFAN movement for the groups whose MFAN node is the first node in each level. At checkpoint 1, we perform MFAN movement for those with MFAN node being the second node in each level. Essentially, the MFAN node ID that we need to perform MFAN movement can be computed by taking checkpoint modulo the number of nodes in each MFAN level. This will give us MFAN (0, 0, 2) for checkpoint 2, which corresponds to nodes (0, 1, 5), respectively;  and (0, 1, 3) for checkpoint 3, which corresponds to nodes (0, 2, 6), respectively. 

While this looks good enough, we still observe the similar drawback. That is, at each checkpoint, the corresponding ORAM request will suffer   {$K+1$} more MFAN movements compared with an ordinary ORAM request. We can leverage this by spreading the   {$K+1$} MFAN movements across the $X$ ORAM requests. That is, when each time $Ctr$ hits   {$\frac{X}{K+1}$}, we will perform exactly one MFAN movement for one group. In this way, one ORAM request will see at most one additional MFAN movement operation. 

Finally, we need to choose the appropriate MFAN movement frequency $X$. A small $X$ causes too many MFAN movements and increases the overhead of the wear-leveling algorithm due to additional memory accesses incurred while a very large $X$ causes too few MFAN movements, hindering efficient wear-leveling.
After experimenting with different values of $X$, we observed that for a Wmax of $10^7$~\cite{one_percent_lifetime}, that a frequency of 10000 incurs about 0.1\% increase memory accesses while also helping us achieve significant lifetime improvements. The detailed performance impact of $X$ will be discussed in \ref{sec:performance_overhead}.

\section{Algorithm Overhead}
\label{sec:algorithm_overhead}
In this section, we discuss the performance and storage overheads incurred by \DESIGN.

\subsection{Performance Overhead}
\label{sec:performance_overhead}
The performance overhead incurred by \DESIGN can be broken down into two components: (i) overhead due to additional memory reads and writes incurred for MFAN movements and (ii) overhead due to additional logic in the ORAM controller for logical node ID to physical node ID translation, etc.

\textbf{MFAN movement overhead:} Systems with ORAM are usually bottlenecked by memory due to the sheer number of memory accesses that need to be performed for each ORAM access. So, calculating the number of additional memory accesses performed due to \DESIGN gives us a reasonable approximation of the performance overhead incurred.

For each MFAN movement, a total of two node reads and two node writes need to be performed. From our node remapping algorithm in section \ref{sec:node_remapping}, we know that an MFAN movement occurs every   {$\frac{X}{K+1}$} ORAM accesses, where $X$ is the wear-leveling frequency and   {$K$} is the MFAN level.
The total number of node reads (or writes) performed for every   {$\frac{X}{K+1}$} ORAM requests (not counting the additional number of reads or writes caused by wear-leveling) is   {$\frac{XL}{K+1}$} and our algorithm introduces an additional two reads (or writes), bringing the overhead to:
\begin{equation}
    MFAN_{overhead} = \frac{2}{\frac{XL}{K+1} } = \frac{2(K+1)}{XL}
\end{equation}

For a 32-level ORAM tree, its MFAN level is $27$, and \DESIGN would cause a 0.0175\% increase in the number of reads and writes, a negligible overhead.

\ignore{It is clear that the performance overhead is composed of the time needed to perform memory swaps, as well as the time needed to compute the translation from logical node ID to physical node ID.} 

\ignore{Let's first look at the MFAN movement overhead. It is obvious that the MFAN movement overhead is dominated by the additional number of memory accesses. Fortunately, with the very simple algorithm \ref{algo:new_wl_algorithm}, it is very clear that an MFAN movement will occur every $\frac{X}{r}$ ORAM accesses. Each MFAN movement will incur 2 memory reads and 2 memory writes, therefore, the MFAN movement overhead is simply}

\ignore{\begin{equation}
    MFAN_{overhead} = \frac{4}{2\frac{X}{K+1} L} = \frac{2(K+1)}{XL}
\end{equation}}
\ignore{The $K+1$ term is used because we need to number of MFAN levels, instead of the ID of the highest MFAN level. 
Notice that this equation assumes the time taken to perform read and write is the same, which is not necessarily true in the real world. However, it does not affect the correctness of the conclusion. If we plug in the parameters used in Fig. \ref{fig:partition_example}, we will see that the overhead is essentially $\frac{2*(2+1)}{10000 * 5} = 0.012\%$. Even for a very large ORAM tree with 32 levels, where $K = 27$, it is expected to achieve $ 0.0175\%$ overhead, which is almost negligible.}

\textbf{ORAM controller overhead:} The other performance overhead is from additional operations performed at the ORAM controller, like lookup table reconstruction, logical node ID to physical node ID translation, etc. We did RTL level synthesis of such hardware, which  contains combinational logic, and the result shows that the overhead is small and can be parallelized with other ORAM controller actions on the critical path of an ORAM access.
Hence this part of the performance overhead is also negligible.

\subsection{Storage Overhead}
\label{sec:storage_overhead}
The storage overhead incurred by our algorithm is entirely on-chip for: (i) a counter to track the number of ORAM accesses performed and (ii) a lookup table for storing the static partition information, which is used to compute node remapping in \ref{sec:node_remapping}. For the counter, a 64-bit counter is enough to track $2^{64}$ ORAM accesses, well over the lifetime of a 32-level ORAM tree even with a Wmax of $10^8$.

\ignore{Essentially, the storage overhead is purely on-chip, because on the PM side, we do not reserve any empty nodes, which is used in Start-Gap \cite{Start-Gap}. There is no storage overhead caused from the PM side. On the CPU side, the overhead is caused by a bunch of variables: the counter $Ctr$ to store the number of ORAM accesses so far, as well as the lookup table used to store the partition information.} 

\ignore{The counter needs to be stored in the persistent domain, and we need a sufficiently large counter to store its value. So how many bits should we give to counter? Suppose we have an L level ORAM tree, and each node tolerates Wmax writes before failing. Then, since $Ctr$ counts the number of ORAM writes, then as long as we ensure $Ctr$ is less than the maximum number of ORAM requests before device fails, we should be fine. Certainly, the number of ORAM requests under ideal   {wear-leveling} is $\frac{Wmax \times (2^L-1)} {L} \approx \frac{Wmax \times 2^L}{L}$. The number of bits needed to store counter is the log2 of the above number. If we take Wmax equals $10^8$, which is a pretty conservative estimation, and L equals 32, then we find that we need 54 bits to store the counter. This implies that a 64 bit counter is sufficiently large to store the number of ORAM requests, and it is able to scale linearly as Wmax and PM capacity scale exponentially due to technology advances. In other words, it is sufficient if we preserve a 64-bit on-chip counter in the persistent domain.} 

The other storage overhead is the lookup table where each entry contains four elements: one Boolean variable 'MFAN level', and 3 integers used to store level information. The Boolean variable occupies 1 bit per entry. For the three integers, 6 bits each is enough for trees with up to 64 levels. Hence, the lookup table overhead is
\ignore{we do not necessarily need 32 bits per integer, because they represent level information. A 6 bit integer is sufficient to represent 64 levels, which is large enough in the foreseeable future. People are unlikely to use an ORAM tree with a level higher than 64.}
\begin{equation}
LookupTable_{overhead} = (6 \times 3+1) \times L  = 19L \mbox{(bits)}
\end{equation}
For a 32-level ORAM tree, the lookup table overhead is merely 76 bytes, leading to 84 bytes overhead in total. This concludes that \DESIGN incurs minimal storage overhead.

\section{Discussion}
\label{sec:discussion}

{\textbf{Security implications:}} While our algorithm significantly improves  NVM lifetime, it breaks away from ORAM algorithm in that under ORAM, every program memory access needs to be translated into an ORAM access. In \DESIGN, when an MFAN movement is triggered, we do not further translate this into ORAM accesses. We argue that this change does not break ORAM's obliviousness nor does it add any new security vulnerabilities. This is so, as with this, the attacker now only additionally knows that an MFAN movement is triggered, and   {$\frac{X}{K+1}$} ORAM accesses have been executed since last MFAN movement. This does not leak information regarding the application unique behavior apart from the number of ORAM accesses that have been performed, which the attacker is already privy to. 

We will now discuss how \DESIGN could be potentially integrated with existing variations of ORAM that aim to improve ORAM performance.

  {\textbf{Ring ORAM:}} { Ring ORAM~\cite{ring_oram} saves ORAM bandwidth by fetching a single block per path read, instead of the entire path, and writing back a deterministic path every $A$ path read operations. Every path read operation will incur an update to that block metadata. This significantly saves ORAM bandwidth and extends NVM lifetime (measured in days) by $A$ times. However, Ring ORAM still incurs skewed access pattern, and E-ORAM could work well with it. Further optimizations could be made if we incorporate E-ORAM with Ring ORAM, but this is out of the scope of this paper. }

\textbf{DRAM+NVM \cite{PCM_as_second_storage}:} Qureshi et al.~\cite{PCM_as_second_storage} proposed a DRAM and NVM hybrid memory system. If we apply ORAM on this hybrid memory system, we could 1) cache some blocks on smaller tree on the DRAM (called $\rho$), and store all blocks on the NVM, (This architecture is proposed in \cite{rho}.) or 2) split the address space across both DRAM and NVM, such that the top more frequently written nodes are mapped to DRAM. For the first scenario, the lifetime of the NVM could be extended by receiving less frequently writes, but since the tree size of the NVM does not change, and the access pattern to the NVM does not change, so the lifetime (percentage lifetime) of the NVM does not change.
For the second scenario, the DRAM acts as a bigger tree-top
cache \cite{phantom}. If the DRAM size is 3\% of NVM size
\cite{PCM_as_second_storage}, and the ORAM tree in the NVM
has 32 levels (NVM tree), then the ORAM tree in the DRAM (DRAM tree) has 27 levels. This extremely large tree-top cache leads to the access pattern to the NVM to be $2^{27}$ smaller ORAM tree with five levels each. Admittedly, if we apply baseline start-gap on these $2^{27}$ 5-level ORAM trees, the lifetime could be significantly boosted without E-ORAM. However, we claim that the performance of setup 1) outperforms setup 2). If we define the DRAM access latency as $T_D$, and define NVM access latency as $T_N$, then we could compute the average memory access time (AMAT) for both setups:
\begin{eqnarray*}
AMAT_1 & = & 27T_D + r_m\times 32T_N \\
AMAT_2 & = & 27T_D + 5T_N 
\end{eqnarray*}
where $r_m$ refers to the $\rho$ miss rate. It is not hard to show that as long as $r_m$ is less than $\frac{5}{32} = 15.6\%$, $AMAT_1$ will be less than $AMAT_2$. 

Since Setup 1) could give user better performance, and its NVM has a bigger tree, then E-ORAM will certainly outperforms Start-Gap as being measured by lifetime. 

\begin{table}[htp]
\centering
 \begin{tabular}{||c | c ||} 
 \hline 
  \multicolumn{2}{||c||}{\textbf{Processor}}        \\  [1ex]  \hline
 ISA                        & UltraSPARC III ISA    \\ 
 CMP Size and Core Freq.    & 1-core, 2.6 GHz       \\ 
 Re-Order-Buffer            & 128 entry             \\ \hline \hline
 \multicolumn{2}{||c||}{\textbf{ NVM Parameters}}   
 \\  
 [1ex] 
 \hline
 \multirow{3}{*}{  NVM Device Parameters}   
                                            & 2 ranks per channel,  \\
                                            & 8 banks per chip,     \\
                                            & 32768 rows per bank   \\ 
  NVM Bus Frequency         & 2666 MHz               \\ 
  NVM Write Queue Size      & 64 entries            \\ 
  NVM Cache Line Endurance & $10^7$ writes         \\ \hline\hline
\multicolumn{2}{||c||}{\textbf{ORAM Configuration}} \\  [1ex] \hline
 Block Size                 & 64B              \\ 
 Bucket Size                & 4 cache lines        \\ 
 ORAM Tree Level            & 24                    \\  
 PLB Size                   & 64KB                 \\ 
 Encryption/Decryption Latency & 21 cycles          \\ 
 Number of Recursive PosMaps & 4                    \\ 
 Stash Size                 & 200 entries           \\ \hline 
\end{tabular}
 \caption{\textbf{Simulated System configuration}}
\label{table:CPU_config}
\end{table}

\ignore{
\begin{figure} 
     \centering
     \subfloat[Slow down caused by \DESIGN compared to no {wear-leveling} for {SPEC CPU\textsuperscript{\textregistered} 2017} benchmark.]{\includegraphics[width=0.9\columnwidth]{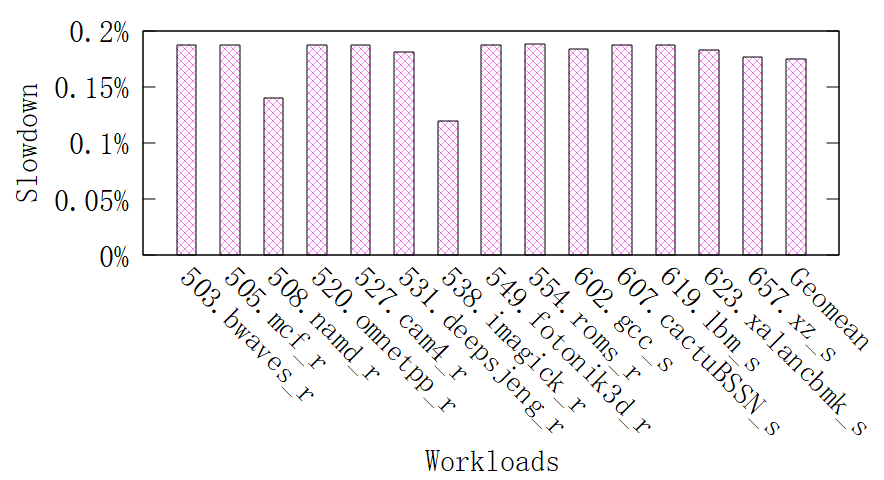}\label{fig:slow_down_spec}}%
    \linebreak
    \centering
     \subfloat[Slow down caused by \DESIGN compared to no {wear-leveling} for GAP benchmark.]{\includegraphics[width=0.5\linewidth]{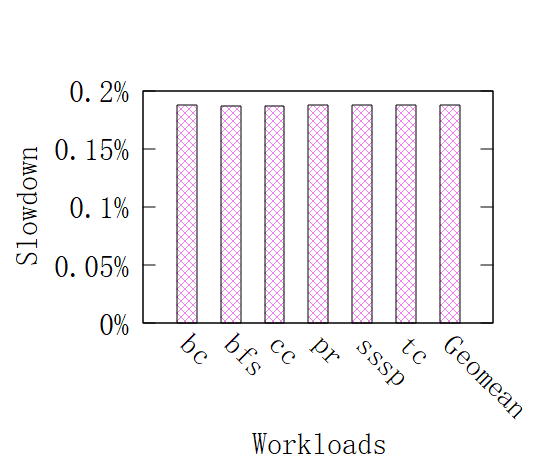}\label{fig:slow_down_gap}}%
        \caption{\textbf{Performance overheads due to \DESIGN} This figure shows the performance overheads for memory intensive workloads from {SPEC CPU\textsuperscript{\textregistered}} 2017 and GAP Benchmarks. The main takeaway is that \DESIGN introduces negligible performance overhead, about 0.2\% when compared to a baseline of no {wear-leveling}, across all benchmarks.}
        \label{fig:slow_down}
\end{figure}
}


 \begin{figure}[t]
    \centering   \includegraphics[width=0.6\linewidth,height=10em]{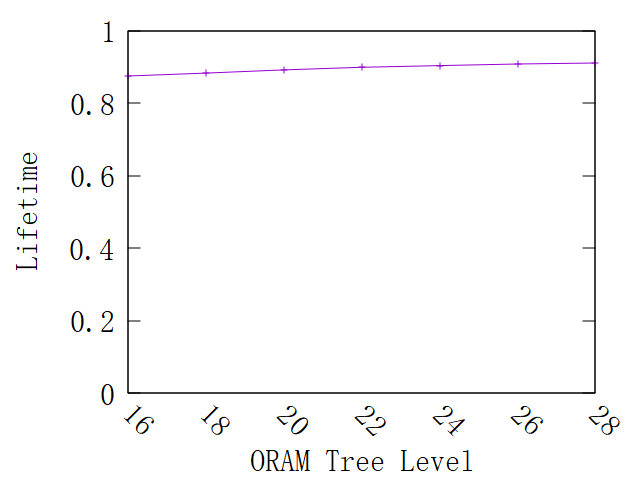}
    \caption{\textbf{ Simulated NVM lifetime with \DESIGN for different tree sizes.} \DESIGN achieves 90\% or higher of the ideal lifetime for different trees sizes versus 4\% with Start-Gap~\cite{Start-Gap}.}
    \label{fig:lifetime_result_tree_size}
\end{figure}
 \begin{figure}[t]
    \centering
    \includegraphics[width=0.9\linewidth,height=10em]{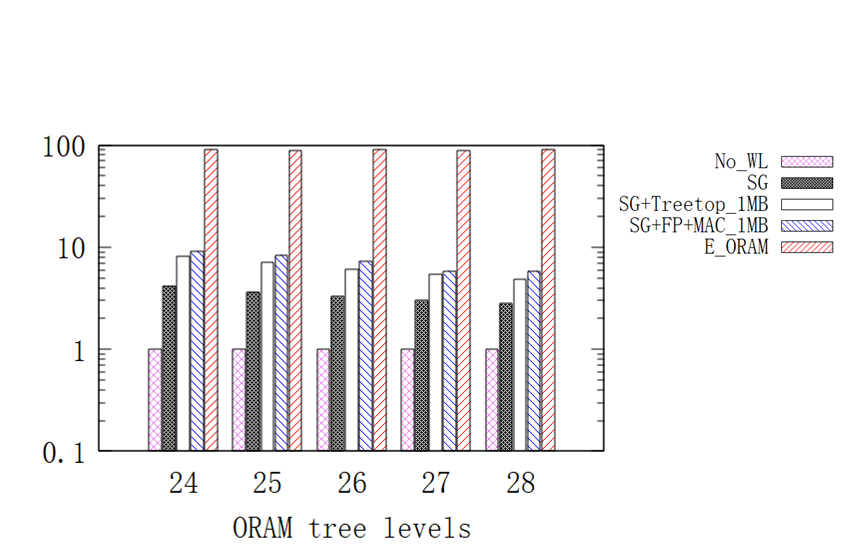}
    \caption{\textbf{\DESIGN vs state-of-the-art:} This figure shows the expected number of ORAM accesses before device failure normalized to baseline no-wear-leveling case, for various state-of-the-art ORAM and wear-leveling configurations and ORAM trees. The results are plotted in log scale, with a value of 100 being ideal. \DESIGN achieves near ideal lifetime while the state-of-the-art is at least 10x worse.}
    \label{fig:lifetime_comparison}
\end{figure}

 \begin{figure}[t]
    \centering   \includegraphics[width=0.9\columnwidth]{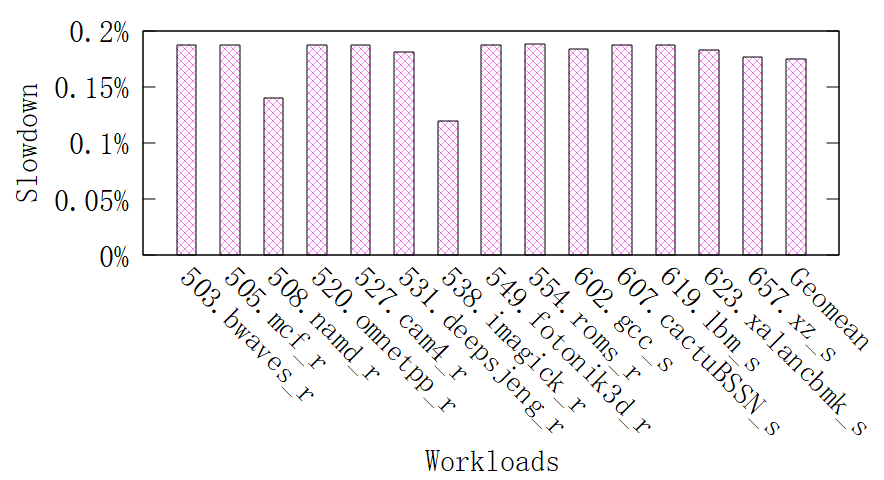}
    \caption{\textbf{\DESIGN performance overhead:} \DESIGN introduces negligible slowdowns ($\leq$ 0.2\%) versus no wear-leveling, across memory intensive {SPEC CPU\textsuperscript{\textregistered} 2017} workloads.}
    \label{fig:slow_down}
\end{figure}

\section{Evaluation}
\label{sec:evaluation}

\subsection{Methodology}
\label{sec:methodology}
We evaluate the performance overhead of \DESIGN by applying USIMM, a  trace-based cycle accurate memory simulator \cite{usimm}. We implemented Freecursive ORAM \cite{freecursive} as well as our wear-leveling engine on USIMM. On the backend, we employed an FR-FCFS memory scheduling policy \cite{fcfs}.
For the ORAM tree, we simulated a relatively small 24-level tree, due to very long simulation time for larger trees.
Table \ref{table:CPU_config} shows the configuration of our simulator and the ORAM tree. For traces, we employed a similar mechanism with \cite{compact}. We select 15 workloads from state-of-art benchmarks, 15 from SPEC CPU\textsuperscript{\textregistered} 2017\cite{spec_2017_benchmark} \footnote{More information about SPEC CPU\textsuperscript{\textregistered} 2017 can be obtained from https://www.spec.org/cpu2017.}
\footnote{SPEC CPU\textsuperscript{\textregistered} is a registered trademark of the Standard Performance Evaluation Corporation.}. These workloads are selected for the fact that these workloads are the most memory-intensive workloads \cite{memory_intensive_workloads, compact}. The traces are then collected for 5 million memory reads and writes, after workloads being fast-forwarded to warm up the LLC. 

\textbf{State-of-the-art comparisons:} We compare \DESIGN with five ORAM settings/configurations: 1) ORAM algorithm without any {wear-leveling} algorithm nor any variants of ORAM, which is our baseline; 2) ORAM algorithm {accommodated} with Start-Gap \cite{Start-Gap}; 3) ORAM algorithm {accommodated} with Start-Gap and a 1MB treetop cache \cite{phantom}; and 4) Fork Path algorithm \cite{fork_path} augmented with Start-Gap and a 1MB merge-aware-cache (MAC). For Start-Gap, we use 256 equal-size groups to maximize lifetime. 

\subsection{Results}
\label{sec:evaluation_results}

\textbf{ NVM lifetime for varying ORAM tree sizes:} Fig. \ref{fig:lifetime_result_tree_size} shows the simulated lifetime for different ORAM tree sizes. As shown in Fig. \ref{fig:lifetime_result_tree_size}, \DESIGN achieves a 87.45\% lifetime for a 16-level ORAM tree, and the lifetime gradually increases as we scale the ORAM tree. The lifetime is able to reach 91.04\% for a 28-level tree. We are not able to simulate a bigger ORAM tree because of the very long simulation time. We can induce that as the  NVM size grows, lifetime continues to grow.

\textbf{Comparison with the state-of-the-art:} Fig. \ref{fig:lifetime_comparison} shows the simulation results when comparing \DESIGN with other four comparison designs mentioned in \S\ref{sec:methodology}.
We simulate the {five} configurations with different ORAM tree {sizes and} collect the number of ORAM accesses being done before device fails. The y-axis is computed by normalizing the total number of ORAM accesses collected from a configuration with the number collected from the baseline configuration. We plot the result in log scale because we observe a huge lifetime gain from \DESIGN. As shown in Fig. \ref{fig:lifetime_comparison}, the baseline configuration has a number of 1. Start gap performs roughly 4x  better compared to the baseline. With the help of 1MB treetop cache, start gap could further improve the number of ORAM accesses to around 8x. If we apply Fork Path on top of start gap, and employ a MAC, due to Fork Path's nature of merging redundant node accesses, we could further improve the number of ORAM accesses to 9x maximum. Compared to {all} these configurations, \DESIGN could significantly boost the number of ORAM accesses to 90x. This is even 10x compared to configuration 4), the Fork Path configuration. The reason that \DESIGN works so well is that \DESIGN dynamically adjust the grouping based on the expected number of writes per node. This information is embedded in the nature of {ORAM and} is not employed by previous {wear-leveling} algorithms. 

\textbf{Performance impact:} Fig. \ref{fig:slow_down} shows the slowdown caused by \DESIGN for {SPEC CPU\textsuperscript{\textregistered}} 2017 benchmarks. As shown in Fig. \ref{fig:slow_down},
\DESIGN incurs negligible performance overhead, ranging from 0.119\% (538.imagick\_r) to 0.187\% (554.roms\_r). 
{We observed similar results with GAP benchmarks} \cite{gap_benchmark}. {We also measured that additional memory accesses generated by} \DESIGN {are about 0.22\%.}

\section{Related Work}
\label{sec:related}

\textbf{ORAM:} Several works~\cite{oram_improvement_3, freecursive, phantom, fork_path, rho} build and improve upon Path ORAM \cite{path_oram}, considered one of the most state-of-art implementation of ORAM primitive. Our work is orthogonal and complementary to these proposals.
\textbf{{Wear-leveling} and write reduction:} Prior solutions to tackle  NVM endurance either reduce write traffic to  NVM (write reduction) or employ wear-levelling to even out write traffic to memory. Our proposed solution \DESIGN is orthogonal to write reduction techniques such as LEO~\cite{leo} which reduce  NVM writes by decreasing block encryptions. 

Park et al. \cite{age_based_wl} proposed an age-based adaptive swapping and shifting {wear-leveling} algorithm that keeps track of the number of writes. Based on the tracked access pattern, they swap pages and shift lines in a page. However, this design introduces relatively large on-chip storage overhead compared to \DESIGN. Ferreira et al. \cite{random_swap_wl} proposed a {wear-leveling} algorithm that randomly picks {two} pages to be swapped. While this design is simple, it fails to achieve uniform write pattern in the case of ORAM, because the most frequently written node (the root node) is unlikely to be picked due to the large {number} of nodes in the ORAM tree. Hu et al. \cite{software_wl} proposed a {software-based} wear-levelling algorithm to spread data across PCM uniformly. However, this technique cannot be applied in the presence of ORAM, because the ORAM controller will still create an exponentially distributed memory access pattern. Chang et al. \cite{march_based_wl} proposed a wear-levelling technique using sliding window with dynamic window size. Memory writes to addresses within the window are tracked. The memory lines are swapped based on the number of observed writes. 

\ignore{
Qureshi et al. \cite{Start-Gap} design Start-Gap {wear-leveling} algorithm to balance the write distribution on PCMs. Start-Gap divides PCM into equal-sized groups. Each group reserves an empty cache line. The CPU maintains a start register and a gap register for each group. Each memory address will be translated based on the value of start and gap registers. The CPU maintains a counter for each group to store the number of writes to that group so far. Upon the counter reaching a threshold, the gap register will be updated, so one of the cache lines in that group will be mapped to a new address. Once the gap register finishes looping through the entire group, the start register will be updated. Gogte et al. \cite{one_percent_lifetime} proposed using software management system to migrate frequently accessed pages from  NVMs to DRAM.}

\section{Conclusions}
\label{sec:conclusion}
ORAM incurs an exponentially distributed memory write pattern causing significant lifetime reductions in  NVM devices due to their limited write endurance.
We present \DESIGN, that leverages this skewed, but deterministic memory access pattern to mitigate  NVM lifetime reductions.
\DESIGN uses a pre-computed static partition grouping technique to determine {wear-leveling} granularity and an efficient intra-group wear-leveling algorithm. Finally, we show that \DESIGN introduces less than 0.2\% performance overhead, negligible storage overhead (84 bytes for a 32-level ORAM tree), with 90x lifetime boost compared to no {wear-leveling}.




\bibliographystyle{plain}
\bibliography{main}

\end{document}